\documentclass[11pt]{article}
\usepackage{epsfig}

\def\be{\begin{equation}}
\def\ee{\end{equation}}

\newcommand{\ssh}[1]{#1\!\!\!/}

\begin{document}
\pagestyle{empty}
\begin{flushright} UPRF-2001-06\\
\end{flushright}

\begin{center} {\Large\bf The structure of radiatively induced Lorentz and
CPT violation in QED at finite temperature}

L. Cervi, L. Griguolo\\ Dipartimento di Fisica, Universit\'a di Parma and
INFN, Gruppo Collegato di Parma,\\ Viale delle Scienze, 43100
Parma, Italy
\\ D. Seminara
\\ Dipartimento di Fisica, Universit\'a di Firenze and INFN,
Sezione di Firenze,\\
Largo Enrico Fermi, 50125 Firenze, Italy
\end{center}

\begin{abstract}

We obtain the induced Lorentz- and CPT-violating term in QED at finite
temperature using imaginary-time formalism and dimensional regularization.
Its form resembles a Chern-Simons-like structure, but, unexpectedly,
it does not depend on the temporal component of the fixed $b_\mu$ constant
vector that is coupled to the axial current. Nevertheless Ward identities are
respected and its coefficient vanishes at $T=0$, consistently with previous
computations with the same regularization procedure, and it is a non-trivial
function of temperature. We argue that at finite $T$ a Chern-Simons-like
Lorentz- and CPT-violating term is generically present, the value of its
coefficient being unambiguously determined up to a $T-$independent constant,
related to the zero-temperature renormalization conditions.
\end{abstract}

\newpage
\pagestyle{plain}
\section{Introduction}
%The possibility of spontaneous breaking of Lorentz- and CPT- invariance at
%low energy has been considered in recent years: possible signals of
%Lorentz- and CPT-violation could be therefore indicative of new physics,
%e.g.. Phenomenological
%consequences of breaking Lorentz- and CPT- symmetry in QED were studied some
%time ago in
% while more recently  possible
% In particular, from a
 %field
%theoretical point view, there arose a controversy on the appearance of a
%Chern-Simons like term generated through radiative corrections \cite{JK}.
The phenomenological consequences of breaking Lorentz- and CPT-invariance
have been actively explored in recent years as they could be measurable
low-energy effects of quantum gravity \cite{QG} or superstrings \cite{Str}.
In QED this issue was examined some time ago in Ref. \cite{CFJ}, while lately
CPT and Lorentz non-invariant extensions
of the Standard Model were scrutinized in Ref. \cite{Colla}.
As many breaking terms are allowed, most efforts have been focused on the
possible  constraints coming from experimental data \cite{Astro}
as well as from renormalizability requirements and anomaly cancellation.
In this context, there arose a ``theoretical'' controversy on
the possibility of generating, through radiative corrections, a
Chern-Simons like term in the effective action of QED. There Lorentz
and CTP symmetries can be in fact destroyed by considering a term of the form
\be
{\cal L}_ {CS}= \frac{k_\mu}{2}\epsilon^{\mu\nu\alpha\beta}A_\nu
 F_{\alpha\beta},
\label{Chern}
\ee
where $k_\mu$ is a constant vector. This breaking term,
suggested in Ref. \cite{CFJ}, predicts birifrangence of the light in the
vacuum and observations on distant galaxies put a very stringent bound on
$k_\mu$ \cite{Astro}. On the other hand the superstring inspired extensions
of the Standard Model proposed in Ref. \cite{Colla} contain,
in the fermionic sector, a Lorentz- and CPT-violating axial-vector coupling
\be
{\cal L}_b=b_\mu\bar{\psi}\gamma^\mu\gamma_5\psi, \label{Axial}
\ee with $b_\mu$ a constant, prescribed four-vector that couples
to the usual axial-current of QED. The interaction term ${\cal
L}_b$ could generate, through radiative corrections, a
non-vanishing value for $k_\mu$ \cite{JK}. If this were the case,
the strong bounds on $k_\mu$ would translate into strong bounds
for the non-invariant term (\ref{Axial}). The aforementioned
controversy arises from the fact the calculation is plagued by a
dependence on the regularization adopted. While some papers
 \cite{chan} claim that particular methods offer the correct result,
others argue that the requirement of vector gauge invariance
forces a vanishing induced term \cite{CG,Bono}. Recently this issue 
was also discussed in \cite{Sit} in the heat kernel approach.
A rather lucid
discussion of the problem appeared in Ref. \cite{Pere}, where it
was pointed out that the relevant form of the vectorial Ward
identities may depend on how the vector $b_\mu$ is embedded into
(or derived from) a more fundamental theory. As an example of that
in \cite{Pere} it was proposed an axion-like model to generate
$b_\mu$ as VEV of a dynamical field: there, a weaker form of the
vectorial Ward identity governs the appearance of the interaction
(\ref{Chern}) in the effective action, and, in particular, it is
not strong enough to ensure the vanishing of its coefficient. In
any case, when  $b_\mu$ is considered a strictly constant
non-dynamical vector field, only vectorial Ward identities with
vanishing axial momentum are relevant and they do not fix the
actual value of the coefficient of the Chern-Simons term: it
depends on the renormalization condition.

Our interest is instead devoted to a different feature of the
problem: the purpose of this letter is in fact to study the effect
of a thermal bath on the structure of the Chern-Simons like term
(\ref{Chern}), obtained by integrating out fermions coupled to the
axial-vector $b_\mu$. In particular our starting observation is
that regularization ambiguities cannot modify temperature
dependence, since they are related to the ultraviolet behavior of
the theory, that is temperature independent. Renormalization
conditions are usually implemented at $T=0$, where the parameters
of the theory are defined, and consequently their temperature
evolution is determined. Our one-loop computation may suffer,
therefore, of the mentioned $T=0$ ambiguities while the functional
form of the induced term and the temperature dependence of its
coefficient are safe. We will use imaginary-time formalism and,
for simplicity, dimensional regularization: in this scheme, where
the vectorial Ward identities hold even at non-zero axial-vector
momentum, a consistency check of our algebra is given by the
vanishing of the induced term  (\ref{Chern}) in the limit
$T\rightarrow 0$. The fact that dimensional regularization in its
standard form does not allow the appearance of the CS term at
$T=0$ has been already pointed out in \cite{Bono}. On the contrary
a CPT- and Lorentz-violating Chern-Simons action {\it is}
generically present at $T\neq 0$: while this fact may have some
relevance for phenomenological application, potentially being
active in the early universe, a serious question arises about the
consistency of the effective theory. For time-like $b_\mu$ it was
shown in \cite{AS} that the vacuum is unstable under pairs
creation of tachyonic photon modes with finite vacuum decay rates
and, recently, it was argued \cite{ada} that, in this case,
unitarity itself may be in trouble (see also the original
discussion in \cite{CFJ}). A more general analysis on the
consistency of the theory at quantum level has been presented in
\cite{leh}, where both time-like and space-like cases appear to be
problematic when microcausality and stability are examined. Rather
surprisingly our computations show that the induced term does not
depend on the temporal component of $b_\mu$: we have not explored
up to now the dynamical consequences of this fact in our
finite-temperature context. Moreover invariance under small (i.e.
not wrapping around the compactified imaginary time  \cite{DGS})
vectorial gauge transformations of the induced term is easily
shown due to the use of dimensional regularization.

\section{The structure of one-loop self-energy at $T\neq 0$}
To begin with let us consider a modified QED action described by the
Lagrangian density
\be
{\cal L}=\bar{\psi}[i\ssh{\partial}-m-\gamma_5\ssh{b}-e\ssh{A}]\psi.
\label{AxialQED}
\ee
As discussed in Ref. \cite{JK} the $b_\mu$ linear contribution to the
 Chern-Simons term arises
from the photon self-energy with one insertion of the axial-vector field,
\be
\Pi^{\mu\nu}_b(p)=i b^\lambda [I_{\mu\nu\lambda}(p)+I_{\nu\mu\lambda}(-p)],
\ee
where $I_{\mu\nu\lambda}$ is given by the "triangle"-like graph with zero
 momentum axial vertex
(we are working from now on directly in Euclidean space):
\be
I_{\mu\nu\lambda}(p)= -ie^2\int \frac{d^4k}{(2\pi)^4}
\frac{{\rm Tr}[\gamma_\mu(\ssh{k}+im)
\gamma_\lambda\gamma_5(\ssh{k}+im)\gamma_\nu
((\ssh{k}+\not{p})+im)]}{(k^2+m^2)^2((k+p)^2+m^2)}.
\label{Pai}
\ee
The CPT- and Lorentz-violating Chern-Simons action is extracted from Eq.
(\ref{Pai}) by isolating, from the odd-parity part, the tensorial structure
linear in the external momentum and by performing the limit
 $p^2\rightarrow 0$ in the scalar integral multiplying it.
In Ref. \cite{Bono} the explicit
evaluation at $T=0$ of $\Pi^{\mu\nu}_b(p)$ in the dimensional regularization
was presented. In particular it was noticed that the only algebraic
properties of
$\gamma_5$ used in the computation were: a) the trace of $\gamma_5$ with an
odd number of Dirac matrices vanishes and b) the trace of $\gamma_5$ with an
 even number of Dirac matrices can be reduced using the Clifford algebra to
the quantity ${\rm Tr}[\gamma_\mu \gamma_\nu \gamma_\alpha \gamma_\beta
 \gamma_5]$. Consistency also requires that ${\rm Tr}[\gamma_\mu \gamma_\nu
\gamma_5]={\rm Tr}[ \gamma_5]=0$.

In this zero-temperature case, the linear $p_\mu$ dependence is easily
extracted and the result can be presented as
\be
-i\frac{e^2}{8\pi^2}b^\lambda p^\beta{\rm Tr}\left[\gamma_\mu \gamma_\nu
\gamma_\lambda \gamma_\beta \gamma_5\right]
\left[{\cal F}_1(p^2/m^2)+{\cal F}_2(p^2/m^2) \right],
\label{Paiodd}
\ee
where the explicit form of ${\cal F}_1(p^2/m^2)$ and ${\cal F}_2(p^2/m^2)$
is given in \cite{Bono}: evaluating ${\cal F}_1(p^2/m^2)$ and
${\cal F}_2(p^2/m^2)$ in $D$ dimensions, taking the limit $D=4$ and
expanding in $p^2$, it results that
\be
{\cal F}_1(p^2/m^2)+{\cal F}_2(p^2/m^2)\simeq - \frac{p^2}{12 m^2},
\ee
showing the absence of a Chern-Simons contribution to the effective action.
We agree with this computation but we want to recall a couple of
remarks in order to better appreciate the
finite $T$ effects. First of all one can check that the cancellation
of the leading order (constant) contribution to ${\cal F}_1(p^2/m^2)+
{\cal F}_2(p^2/m^2)$ comes from a delicate balance between a
"classical" term (proportional to $m^2$
in Eq.(\ref{Pai})) and an "anomalous"
quantum term, deriving from the potential divergences\footnote{
This mechanism of cancellation of a quantum contribution against a
classical one is reminiscent of the analogous phenomena in $3D$ in
the case of parity anomaly. This perfect balance between the two contributions
is peculiar of the standard dimensional regularization. In other scheme this 
exact cancellation does not occur, leaving us with a non vanishing CS term
whose coefficient is however temperature independent.}.
As mentioned  in Ref.\cite{Pere}, the massless case
escapes this mechanism, therefore dimensional regularization gives there a
non-vanishing result.
This fact is related to the loss of analyticity \cite{CG} of the
"triangle"-like diagram at $m=0$
in the limit of vanishing axial-vector momentum.
In the finite temperature case, where the analyticity properties in the external
 momenta are
usually weaker, this observation suggests the concrete possibility that a non-zero
 result could appear even in the massive case. Secondly, we stress that the
covariance of the momentum integration immediately selects the Chern-Simons
 tensorial structure and the dependence on $p^2$ of the coefficient function at
zero-temperature.
 This is no longer true at finite temperature, as we shall see in a while, due
 to explicit presence of the Matsubara frequencies: in particular the double
limit $p_0\rightarrow 0$ and $p_i\rightarrow 0$ has to be performed very carefully.

Let us assume, from now on that the system is in thermal equilibrium with a
temperature $T=\beta^{-1}$, $\beta$ being interpreted as the radius of the
compactified euclidean time. In this case we may use the Matsubara formalism
 that consists simply in taking $k_0=(n+\frac{1}{2})\frac{2\pi}{\beta}$
(anti-periodic boundary conditions for fermions requires semi-integers
frequencies) and replacing
$\frac{1}{2\pi}\int dk_0=\frac{1}{\beta}\sum_{n}$. The remaining
$\int d^3\hat k$ integral is of course continued to $D$-spatial dimensions.
The trace can be still
performed in full generality and simple algebraic manipulations in the
loop momenta
(not involving shifts or symmetry properties) allows us to write
(\ref{Selfy}) as follows:
\begin{eqnarray}
&&I_{\mu\nu\lambda}(p)=\nonumber\\
&&-4 i\frac{e^2}{\beta}\sum_{n}\int \frac{d^D\hat k}{(2\pi)^D}
\frac{(2k_\mu\epsilon_{\nu\lambda\rho\sigma}p^\rho k^\sigma-2k_\nu
\epsilon_{\mu\lambda\rho\sigma}p^\rho k^\sigma)
-2(k\cdot p)\epsilon_{\mu\nu\lambda\rho}p^\rho+ p^2\epsilon_{\mu\nu\lambda\rho}
(k^\rho-p^\rho)}{(k^2+m^2)^2((k+p)^2+m^2)}+\nonumber\\
&&+4 i\frac{e^2}{\beta}\sum_{n}\int \frac{d^D\hat k}{(2\pi)^D}
\frac{\epsilon_{\mu\nu\lambda\rho}(k^\rho-p^\rho)}{(k^2+m^2)^2}.
\label{Selfy}
\end{eqnarray}
The following step in the computation is to extract the tensorial structure,
 leaving us with the
evaluation of scalar integrals. The second term in Eq.(\ref{Selfy}) is easily
tamed,
both the integral over the spatial components and the series over $k_0$ are
 anti-symmetric in exchanging $k\rightarrow -k$ (we can find a region around
$D=3$ where everything is convergent). It remains therefore
\be
-4 i\frac{e^2}{\beta}\,\epsilon_{\mu\nu\lambda\rho}p^\rho I_0,
\ee
where we have
\begin{equation}
I_0=\sum_{n}\int \frac{d^D\hat k}{(2\pi)^D}
\frac{1}{(\hat{k}^2+k_0^2+m^2)^2},
\end{equation}
that exhibits the Chern-Simons like structure.
Let us discuss now the first contribution. We introduce Feynman parameters in
 order to perform
the integral. To implement translations only on the space components of the
loop momentum we
decompose $k_\mu$ as follows
\be
k_\mu=\hat{k}_\mu+k_0\delta_{0\mu}.
\ee
Shifting $\hat{k}\rightarrow \hat{k}-x\hat{p}$
(where $\hat{p}_\mu$ is defined as above)
in Eq.(\ref{Selfy}) and using the covariance under spacial rotations,
which allows us  to conclude
that
$$\hat{k}_\mu\hat{k}_\nu
\rightarrow\frac{\hat{k}^2}{D}(\delta_{\mu\nu}-\delta_{\mu 0}\delta_{\nu 0}),$$
we arrive to the form
\begin{eqnarray}
I_{\mu\nu\lambda}(p)&=& - 4 i\frac{e^2}{\beta}\left [
\epsilon_{\mu\nu\lambda 0} I_1
+\epsilon_{\mu\nu\lambda\rho}p^\rho(2I_2+I_0)\right. \nonumber\\
&+&(p^\mu\epsilon_{\nu\lambda\rho 0}p^\rho -
p^\nu\epsilon_{\mu\lambda\rho 0}p^{\rho} -
p^2\epsilon_{\mu\nu\lambda 0})I_3\nonumber\\
&+&\left. (\delta_{0\mu}\epsilon_{\nu\lambda\rho 0}p^\rho -\delta_{0\nu}
\epsilon_{\mu\lambda\rho 0}p^\rho- p_0\epsilon_{\mu\nu\lambda 0})I_4\right ],
\label{Master}
\end{eqnarray}
where
\begin{eqnarray}
I_1&=& \int_0^1 dx\,2(1-x)\sum_{n}\int \frac{d^D\hat k}{(2\pi)^D}
\frac{[p^2(1-2x)(k_0+xp_0)-\frac{2}{D}\hat{k}^2 p_0+2p_0(k_0+xp_0)^2]
}{[\hat{k}^2+(k_0+xp_0)^2+x(1-x)p^2+m^2]^3}.
 \nonumber\\
I_2&=&\int_0^1 dx\,2(1-x)\sum_{n}\int \frac{d^D\hat k}{(2\pi)^D}
\frac{[-p_0(k_0+xp_0)-\frac{2}{D}\hat{k}^2-\frac{1}{2}p^2(1-x)]}
{[\hat{k}^2+(k_0+xp_0)^2+x(1-x)p^2+m^2]^3}. \nonumber\\
I_3&=&- \int_0^1 dx\,2(1-x)\sum_{n}\int \frac{d^D\hat k}{(2\pi)^D}
\frac{2x(k_0+xp_0)}{[\hat{k}^2+(k_0+xp_0)^2+x(1-x)p^2+m^2]^3}.\nonumber\\
I_4&=&\int_0^1 dx\,2(1-x)\sum_{n}\int \frac{d^D\hat k}{(2\pi)^D}
\frac{(2(k_0+xp_0)^2-\frac{2}{D}\hat{k}^2)}{[\hat{k}^2+
(k_0+xp_0)^2+x(1-x)p^2+m^2]^3}
\label{Is}
\end{eqnarray}
Working at finite temperature, more structures have been generated, 
a fact that is not unexpected due to the explicit breaking of 
four-dimensional covariance.
The important point is that, nevertheless, the tensors must be transverse
respect $p_\mu$ and $p_\nu$, since the vectorial Ward identity are unaffected 
by
the presence of the temperature. By inspection we see that the only potential 
trouble comes
from $I_1$ (being $\epsilon_{\mu\nu\lambda 0}$ not transverse!). Luckily we can
 show that $I_1$ is exactly zero. To this purpose, it is useful to
rewrite $I_1$ as follows
\begin{eqnarray}
I_1&=&\int_0^1dx\,2(1-x)\sum_{n}\int \frac{d^D\hat k}{(2\pi)^D}
\left[- \frac{k_0+xp_0}{2}\frac{d}{dx}
\left[\frac{1}{[\hat{k}^2+(k_0+xp_0)^2+x(1-x)p^2+m^2]^2}\right] \right].\nonumber\\
&-&
\left.\frac{2 p_0\hat{k}^2/D}{[\hat{k}^2+(k_0+xp_0)^2+x(1
-x)p^2+m^2]^3}\right],
\label{I1}
\end{eqnarray}
and integrating by part with respect to $x$ (the boundary terms are zero) we get, after
 having performed the $D-$dimensional integral,
\begin{equation}
I_1=-\frac{\Gamma(2-\frac{D}{2})}{(4\pi)^{\frac{D}{2}}}\int_0^1dx
\sum_{n}\frac{(k_0+xp_0)}{[(k_0+xp_0)^2+x(1-x)p^2+m^2]^{2-\frac{D}{2}}}.
\label{I11}
\end{equation}
We can use now the explicit form of the Matsubara frequencies and the fact
 that $p_0$ is
discrete ($p_0=\frac{2\pi}{\beta}l$): relabeling the sum in Eq.(\ref{I11})
as $n\rightarrow-n-1-l$ and making the change of variables $y=1-x$, one easily
 obtains that
$$I_1=-I_1.$$
We stress that we did our computations in $D$ dimensions, where everything is
convergent and no
limit on $p$ has been performed. The same arguments applies to $I_3$, and we
 remain, therefore,
with two independent tensorial structures: we need to evaluate $I_0+2I_2$
and $I_4$.
Before entering the computations we remark that
the emergence of a new, transverse tensorial structure was overlooked in
Ref. \cite{bra}, where the coefficient of the Chern-Simons term at finite $T$
was obtained simply evaluating the scalar integral, relevant at $T=0$, by
introducing Matsubara frequencies for $p_0$. As we will see in the next
section, our result disagrees with that.
\section{The CPT- and Lorentz-violating term at $T\neq 0$}

Let us evaluate the coefficients of the two independent structures, in
the small momentum limit: we remark that at finite temperature this
procedure is rather delicate, due to the fact that, in general,
the limits $p_0\rightarrow 0$ and $\hat{p}^2\rightarrow 0$ do not
commute  \cite{gross}. Here we shall take first $p_0\rightarrow 0$ and send
$\hat{p}^2\rightarrow 0$. The sum
$I_0+2I_2$ becomes, when the D-dimensional integrals have been computed:
\begin{equation}
I_0+2I_2=-\frac{\Gamma\left(2-\frac{D}{2}\right)}{(4\pi)^{\frac{D}{2}}}
\int_0^1 dx 2 (1-x)\sum_{n}\left[\left(1-\frac{4}{D}\right )
\frac{1}{[k_0^2+m^2]^{2-\frac{D}{2}}}+\frac{4}{D}\frac{2-D/2}{\Gamma(3)}
\frac{1}{[k_0^2+m^2]^{2-\frac{D}{2}}}\right].
\label{I12}
\end{equation}
The result is zero ${\it identically}$ in arbitrary dimension: one can check
that when the dependence on $\hat{p}^2$ is retained the corrections are
regular, and of order $\hat{p}^2$ (and of course the coefficient
depends on $T$). We see that the the zero-temperature
tensorial structure still has a vanishing coefficient in the small
momentum limit when $T\neq 0$. The only possible contribution to
Lorentz- and CPT-violation could therefore arise from $I_4$, i.e.
from the non-covariant structure. The relevant term to be calculated is
\be
I_4=\int_0^1 dx 2 (1-x)\sum_{n}\int \frac{d^D\hat k}{(2\pi)^D}
\frac{(2k_0^2-\frac{2}{D}\hat{k}^2)}
{[\hat{k}^2+k_0^2+m^2]^3}.
\label{I44}
\ee
The D-dimensional integration leads to
\begin{equation}
I_4=\frac{\Gamma(2-\frac{D}{2})}{(4\pi)^{\frac{D}{2}}}
\int_0^1dx  (1-x)\sum_{n}\left[(3-D)
\frac{1}{[k_0^2+m^2]^{2-\frac{D}{2}}}+(D-4)
\frac{m^2}{[k_0^2+m^2]^{3-\frac{D}{2}}}\right].
\label{I176}
\end{equation}
At this point we need an explicit representation for the sum over
the Matsubara frequencies: we use the following result
\cite{ford}, valid when $1/2<\lambda<1$
\be
\sum_n[(n+b)^2+a^2]^{-\lambda}=
\frac{\sqrt{\pi}\Gamma(\lambda-1/2)}{\Gamma(\lambda)(a^2)^{\lambda-1/2}}
+4\sin(\pi\lambda)\int_{|a|}^\infty \frac{dz}{(z^2-a^2)^\lambda} {\rm Re}
\left (\frac{1}{\exp 2\pi(z+ib)-1}\right).
\label{Ford}
\ee
We cannot apply directly this formula to our case: at the end we want
to take the $D=3$ limit, and it is clear that, evaluating the second
contribution to $I_4$ in Eq.(\ref{I176}), the integral in
Eq.(\ref{Ford}) do not converge there in the limit $D=3$.
It is not difficult anyway to perform the analytical continuation in
Eq.(\ref{Ford}) using the relation
\begin{eqnarray}
&&\!\!\!\!\!\!\!\!\!\!\!\!\!\!
\int_{|a|}^\infty \frac{dz}{(z^2-a^2)^{\lambda}}
{\rm Re}\left (\frac{1}{\exp 2\pi(z+ib)-1}\right )=\frac{1}{2a^2}
\frac{3-2\lambda}
{1-\lambda}\int_{|a|}^\infty \frac{dz}{(z^2-a^2)^{\lambda-1}}
{\rm Re}\left (\frac{1}{\exp 2\pi(z+ib)-1}\right)\nonumber\\
&-& \frac{1}{4a^2} \frac{1}{(2-\lambda)(1-\lambda)}
\int_{|a|}^\infty \frac{dz}{(z^2-a^2)^{\lambda-2}}
\frac{d^2}{dz^2}\left [{\rm Re}\left (\frac{1}{\exp 2\pi(z+ib)-1}\right)
\right].
\end{eqnarray}
Eq.(\ref{I176}) can now be explicitly evaluated at $D=3$: we see that the
potential singularity at $D=3$, coming from the first contribution,
cancels (notice the factor $D-3$ in front), the finite residue (that
would be temperature independent) cancels with an analogous term coming
from the second contribution, leaving us with the final result
\be
I_4=2\beta \int_{|\xi|}^\infty dz(z^2-\xi^2)^{\frac{1}{2}}
{{\tanh(\pi z)}\over{\cosh^2(\pi z)}}= 2\,\beta\,F(\xi),
\label{Formula}
\ee
where we have defined
$\xi=\frac{\beta m}{2\pi}$. The behavior of $F(\xi)$ is displayed in
the plot below.
\begin{figure}[h]
\begin{center}
\mbox{\epsfig{file=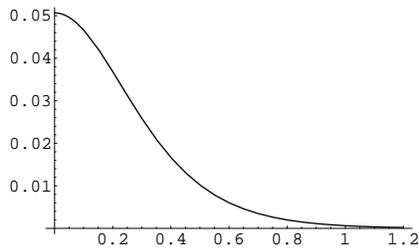,width=5.5cm}}
\caption{plot of $F(\xi)$}
\end{center}
\end{figure}
Eq.(\ref{Formula}) is the main result of our paper: it shows that for
$\beta\neq \infty$ ($T\neq 0$) a CPT- and Lorentz-violating term appears.
In momentum space it can be written as
\be
\Pi^{\mu\nu}_b(p)=4e^2 F(\xi)b^\lambda 
(\delta_{0\mu}\epsilon_{\nu\lambda\rho 0}p^\rho
-\delta_{0\nu} \epsilon_{\mu\lambda\rho 0}p^\rho-
p_0\epsilon_{\mu\nu\lambda 0})+O(p^2). \ee 
Several comments are now in
order: first of all when $T=0$ $F(\xi)$ vanishes, recovering
therefore the fact that, using dimensional regularization, no CPT-
and Lorentz-violating  Chern-Simons like term is present in the
effective action. The opposite limit ($T\rightarrow\infty$) is
otherwise finite ($F(0)=1/2\pi^2$): in Ref. \cite{bra} a similar
behavior was found for the temperature evolution of the
coefficient of the Chern-Simons like term but there the $T=0$
boundary condition was taken so that at $T=\infty$ the symmetries
were restored. Moreover, at variance with our result, the
Chern-Simons term there was implicitly assumed to be related to
the
 ${\it covariant}$ tensorial structure, fact that from our computation
turns out to be incorrect.The second point is that our induced action does
 not depend on the temporal component of $b_\mu$; in refs. \cite{AS,ada,leh}
it was discussed the consistency of the theory at quantum level,
when the Chern-Simons like action is present. It would be interesting to
address the problem of stability in the finite-temperature situation
considering our induced term. Another observation is related
to the dependence on $\xi$: we see that the limit $\beta\rightarrow 0$ is
the same as $m\rightarrow 0 $ (since the only dependence on the mass and on
the temperature appears through $\xi$).
This suggests that the presence at finite
temperature of a non-zero CS-like term is related to the loss of analyticity
in external momenta, bypassing therefore the argument of Coleman and Glashow
\cite{CG} against it (analyticity was also assumed in Ref.\cite{Bono}). It is
interesting to write down the induced term in configuration space
\begin{equation}
S_{``CS"}=4 i e^2 F(\xi)\int d^4x\,b_i\bigl[A_0\epsilon^{ijk}F_{jk}-
2 \epsilon^{ijk}A_jF_{0k}\bigr]
\end{equation}
gauge invariance is achieved via Bianchi identity up to a total derivative.

The asymmetrical behavior played in the above action  by the ``spacial''
and ``temporal''  component of $b_\mu$ (the latter being completely absent)
might seem strange. A source of this asymmetry can be surely traced back to
the presence of a thermal bath which, selecting a specified  frame, provides 
an additional Lorentz violation. However, below, with the help of the 
analogous problem in two dimensions, we would like to suggest that the origin 
of this term at finite temperature  may be related to a deeper geometrical
reason. In D=2, in fact, its  structure and its coefficient  can be easily
understood through the interplay of the  global part of the Quillen anomaly, 
which controls the obstruction to the chiral splitting, and the presence of 
a non trivial cycle (finite temperature in the above language).

There, the relevant Green function is the 1-point function with one
$b_\mu$-insertion (we study the massless case for simplicity). Using
dimensional regularization we see that the zero-temperature computation
gives a vanishing result because $\int d^Dk \frac{k_\mu k_\nu}{k^2}=0$
(we remark that having $b_\mu$ constant implies that the external momentum
 has to be null). But turning on the temperature the situation changes
drastically. The relevant integral is:
\be
\Pi^b_\mu=- i e b^\lambda{\rm Tr}
[\gamma_\mu\gamma_\alpha\gamma_\lambda\gamma_5
\gamma_\nu]
\frac{1}{\beta}\sum_n\int
\frac{d^D\hat{k}}{(2\pi)^D}\frac{k^\alpha k^\nu}{k^4},
\label{due}
\ee
that is equivalent, after using Dirac algebra to
\begin{eqnarray}
&&\,2 e b^\lambda\left[-\epsilon_{\lambda\mu}\frac{1}{\beta}\sum_n
\int\frac{d^D\hat{k}}{(2\pi)^D}\frac{1}{[k_0^2+\hat{k}^2]}
+2\epsilon_{\lambda\alpha}\delta_{0\mu}\delta_{0\alpha}
\frac{1}{\beta}\sum_n\int
\frac{d^D\hat{k}}{(2\pi)^D}{{k_0^2}\over{[\hat{k}^2+k_0^2]^2}}\right.
\nonumber\\
&&\ \ \ \ \ \ \ \ \ \  \ \ \  \left. +2\epsilon_{\lambda\alpha}
\frac{1}{\beta}
\sum_n\int
\frac{d^D\hat{k}}{(2\pi)^D}
\frac{\hat{k}_\mu \hat{k}^{\alpha}}{[k_0^2+\hat{k}^2]^2}\right].
\end{eqnarray}
It is not difficult to see that the $D-$dimensional integration and the
analytical continuation of the sum gives a finite result for the $b_1$
component (due to a cancellation between a pole and a zero as $D=1$)
 leaving us with
\be
\Pi^b_\mu=\frac{16}{\pi} e b^{\lambda}\epsilon_{\lambda\alpha}\delta_{0\mu},
\ee
while one can easily show that $b_0$ component has zero coefficient after
performing the $\hat{k}$-integration. We remark that here no external
momentum limit has been done, therefore the result is exact. The above
term has a natural interpretation in configuration space as
\be
i \int d^2 x b_1 A_0(x), \label{amv} 
\ee 
that is the analogue of
$D=4$ (the ``complete" CS-like  term would be here
$\epsilon^{\mu\nu}b_\mu A_\nu$). The term appearing eq.
(\ref{amv}) is nothing but a remnant of the holomorphic anomaly on
the torus \cite{amv}. In fact, when non-trivial cycles are present
in the base-space, the effective action acquires a subtle
dependence on the harmonic part of the gauge potentials. In
particular when both vector and axial gauge fields are coupled,
requiring gauge invariance implies that an anomalous phase has to
be present in order to cure the transformation properties of the
modulus of the Dirac determinant. This phase can be derived from
general algebraic-geometrical arguments \cite{amv}, being related
to the Quillen anomaly, or by an explicit $\zeta-$function
computation \cite{det} of the relevant functional determinants.
The asymmetric character of the phase has to be ascribed to the
anomalous modular transformation properties of chiral partition
functions. We can now understand the appearance of this term at
finite temperature: $b_\mu$ is  basically an harmonic one-form
axially coupled and therefore being able to interact with the
harmonic component of $A_0$. The complete anomalous phase requires
a quadratic part in $b_\mu$, that can be easily recovered by
computing the Feynman graph with two $b_\mu$ insertions.
This discussion may suggest that the four-dimensional term could have an
appealing mathematical interpretation.

\noindent{\bf Acknowledgements}: We warmly thank Prof. S. Deser,
Prof. R. Jackiw and Prof. V.A. Kostelecky for suggestions and
discussions.


\newpage
\begin{thebibliography}{99}
\bibitem{QG} R.M. Wald, Phys. Rev.{\bf D21}, 2742 (1980); S.W. Hawking,
Phys. Rev.{\bf D32}, 2489 (1985).
\bibitem{Str} V. A. Kostelecky and S. Samuel, Phys. Rev.{\bf D39}, 683 (1989);
V.A. Kostelecky and R. Potting, Nucl. Phys. {\bf B359}, 545 (1991).
\bibitem{CFJ} S.M. Carrol, G.B. Field and R. Jackiw, Phys. Rev.{\bf D41},
1231 (1990).
\bibitem{Colla} D. Colladay and V. A. Kostelecky,
Phys. Rev.{\bf D55}, 6760 (1997), {\bf D58}, 116002 (1998).
\bibitem{Astro}M. Goldhaber and V. Trimble, J. Astrophysics. Astr. 17,
17 (1996); S.M. Carrol and G.B. Field, Phys. Rev. Lett. {\bf 79}, 2394 (1997);
D. Bear et al. Phys. Rev. Lett. {\bf 85}, 5038 (2000).
\bibitem{JK} R. Jackiw and V.A. Kostelecky, Phys. Rev. Lett. {\bf 82}, 3572
(1999).
\bibitem{chan} L. Chan, {\it Induced Lorentz-violating Chern-Simons term in
QED and anomalous contribution to effective action expansion}, HEP-PH/9907349
\bibitem{CG} S. Coleman and S. Glashow, Phys. Lett. {\bf B405}, 249 (1997),
Phys. Rev. {D}59, 116008 (1999).
\bibitem{Bono} G. Bonneau, Nucl. Phys. {\bf B593}, 398 (2001).
\bibitem{Sit} Yu A. Sitenko, {\it One-Loop 
Effective Action for the Extended Spinor Electrodynamics with 
Violation of Lorentz and CPT Symmetry}, HEP-TH/0103215.
\bibitem{Pere} M. Perez-Victoria, {\it Physical (ir)relevance of ambiguities
to Lorentz and CPT violation in QED}, HEP-TH/0102021.
\bibitem{AS} A.A. Andrianov and R. Soldati, Phys. Lett.{\bf B435}, 449 (1998).
\bibitem{ada} C. Adam and F.R. Klinkhamer,
{\it Causality and CPT violation from an abelian Chern-Simons like term},
HEP-PH/0101087.
\bibitem{leh} V. A.  Kostelecky and R. Lehnert, Phys. Rev.{\bf D63},
065008 (2001).
\bibitem{DGS} S. Deser, L. Griguolo and D. Seminara, Phys. Rev.
Lett. {\bf 79}, 1976 (1997), Phys. Rev. {\bf D57}, 7444 (1998).
\bibitem{bra} J.R. Nascimento, R.F. Ribeiro and N.F.Svaiter,
{\it Radiatively induced Lorentz and CPT violation in QED at finite
temperature}, HEP-TH/0012039.
\bibitem{gross} D.J.~Gross, R.D.~Pisarski and L.G.~Yaffe,
Rev. Mod. Phys. {\bf 53}, 43 (1981).
\bibitem{ford}L.H. Ford, Phys. Rev.{\bf D21}, 933 (1980).
\bibitem{amv}L. Alvarez-Gaum\'e, G. Moore and C. Vafa,
Commun. Math. Phys. {\bf 106}, 1 (1986).
\bibitem{det} I. Sachs and A. Wipf, Ann. of Phys. {\bf 249}, 380 (1996),
L. Griguolo and D. Seminara, Nucl. Phys. {\bf B495}, 400 (1997).
\end{thebibliography}
\end{document}